\documentstyle[11pt,aaspp,flushrt]{article}
\catcode`\@=11 
\def\@versim#1#2{\vcenter{\offinterlineskip
        \ialign{$\m@th#1\hfil##\hfil$\crcr#2\crcr\sim\crcr } }}
\newcommand{\beq}{\begin{equation}}
\newcommand{\eeq}{\end{equation}}

\def\lsim{\mathrel{\mathpalette\@versim<}}
\def\gsim{\mathrel{\mathpalette\@versim>}}

\def\mpy{M_\odot \ {\rm yr^{-1}}}
\def\egs{\rm \ ergs \ s^{-1}}
\def\cc{\rm \ cm^{-3}}

\def\rad{\rm \ rad \ m^{-2}}
\def\md{\dot m_{-4}}

\begin{document}
\title{Constraining the Accretion Rate Onto Sagittarius A$^*$ Using
Linear Polarization} \author{Eliot Quataert\altaffilmark{1} and Andrei
Gruzinov} \affil{Institute for Advanced Study, School of Natural
Sciences, Einstein Drive, Princeton, NJ 08540; eliot@ias.edu,
andrei@ias.edu} \altaffiltext{1}{Chandra Fellow}
\centerline{}
\centerline{{\it To appear in the Astrophysical Journal}}
\begin{abstract}
Two possible explanations for the low luminosity of the supermassive
black hole at the center of our galaxy are (1) an accretion rate of
order the canonical Bondi value ($\sim 10^{-5} \ \mpy$), but a very
low radiative efficiency for the accreting gas or (2) an accretion
rate much less than the Bondi rate.  Both models can explain the
broad-band spectrum of the Galactic Center.  We show that they can be
distinguished using the linear polarization of synchrotron radiation.
Accretion at the Bondi rate predicts no linear polarization at any
frequency due to Faraday depolarization.  Low accretion rate models,
on the other hand, have much lower gas densities and magnetic field
strengths close to the black hole; polarization may therefore be
observable at high frequencies.  If confirmed, a recent detection of
linear polarization from Sgr A$^*$ at $\gsim 150$ GHz argues for an
accretion rate $\sim 10^{-8} \ \mpy$, much less than the Bondi rate.
This test can be applied to other low-luminosity galactic nuclei.

\

\noindent {\it Subject Headings:} accretion, accretion disks ---
Galaxy: center --- polarization

\end{abstract} 

\section{Introduction}

In this paper we discuss the effect of Faraday depolarization on
synchrotron radiation in spherical accretion flow models of
low-luminosity galactic nuclei (see also Bower et al 1999ab; Agol
2000).  We focus on the radio source Sgr A$^*$ at the Galactic Center,
but our results can also be applied to other systems (see \S4).  This
paper was motivated by a possible detection of linear polarization
from Sgr A$^*$ (Aitken et al. 2000).

The Bondi accretion rate onto the supermassive black hole at the
center of our galaxy is estimated to be $\sim 10^{-4}-10^{-5} \ \mpy$
(e.g., Coker \& Melia 1997; Quataert, Narayan, \& Reid 1999), implying
a luminosity of $\sim 10^{41} \egs$ if the radiative efficiency is
$\sim 10 \%$.  This is roughly 5 orders of magnitudes larger than the
observed luminosity (see Narayan et al. 1998 for a recent
compilation).  Comparable ``discrepancies'' are obtained for massive
elliptical galaxies in nearby X-ray clusters (e.g., Fabian \& Rees
1995; Di Matteo et al. 1999).

One explanation for the low luminosity of nearby supermassive black
holes is that they accrete via an advection-dominated accretion flow
(ADAF), in which most of the dissipated turbulent energy is stored as
thermal energy rather than being radiated (e.g., Rees et al. 1982;
Narayan \& Yi 1994, 1995; Abramowicz et al. 1995).  In such models the
accretion rate is of order the Bondi rate while the radiative
efficiency is extremely small ($\sim 10^{-6}$ for Sgr A$^*$).

Another explanation for very low luminosity accreting systems is that
the Bondi accretion rate estimate is inapplicable (e.g., Blandford \&
Begelman 1999; Gruzinov 1999).  In particular, numerical simulations
of non-radiating accretion flows with small values of the
dimensionless viscosity parameter $\alpha$ find that the gas density
scales with radius as $\rho \propto r^{-1/2}$ rather than the
canonical Bondi/ADAF scaling of $\rho \propto r^{-3/2}$ (Stone,
Pringle, \& Begelman 1999; Igumenshchev \& Abramowicz 1999, 2000;
Igumenshchev, Abramowicz, \& Narayan 2000).  Narayan, Igumenshchev, \&
Abramowicz (2000) and Quataert \& Gruzinov (2000) explained these
simulations in terms of a ``convection-dominated accretion flow''
(CDAF). In such a flow angular momentum is efficiently transported
inwards by radial convection, nearly canceling the outward transport
by magnetic fields.  This strongly suppresses the accretion of matter
onto the black hole.


Broad band spectra have thus far had difficulty distinguishing between
these explanations for the low luminosity of nearby supermassive black
holes.  For example, Quataert \& Narayan (1999; hereafter QN) showed
that accretion at much less than the Bondi rate could produce spectra
quite similar to ADAF models.  In this paper we show that linear
polarization observations in the radio to sub-mm can provide a
sensitive probe of the accretion rate onto the black hole, and help
distinguish between degenerate spectral models.

In the next section (\S2) we present simple estimates of the physical
parameters of the accretion flow relevant for our analysis.  We then
discuss Faraday depolarization in spherical accretion flow models of
Sgr A$^*$ (\S3).  In \S4 we compare these predictions with
observational constraints on the linear polarization of Sgr A$^*$ and
summarize our results.  We also generalize our analysis to other
low-luminosity galactic nuclei.

Throughout this paper, we focus on accretion models of Sgr A$^*$.  An
unresolved jet or outflow may, however, dominate the observed emission
(e.g., Falcke, Mannheim, \& Biermann 1993; Lo et al. 1998; Falcke
1999); this is briefly discussed in \S4.

\section{Plasma Parameters for Sgr A$^*$}

Stellar kinematics show that there are $\approx 2.6 \times 10^6
M_\odot$ within $\approx 0.015$ pc of the Galactic Center (Eckart \&
Genzel 1997, Ghez et al. 1998), centered on the radio source Sgr A$^*$
(Menten et al. 1997).  The most plausible explanation is that Sgr
A$^*$ is a $\approx 2.6 \times 10^6 M_\odot$ accreting black hole.
Sgr A$^*$ is believed to accrete the winds from nearby ($\sim 0.1$ pc)
massive stars (Krabbe et al. 1991).  The Bondi accretion rate of these
winds onto the supermassive black hole is estimated to be $\approx
10^{-4}-10^{-5} \ \mpy$ (e.g., Coker \& Melia 1997; Quataert, Narayan,
\& Reid 1999).



If the accretion rate close to the black hole is of order the Bondi
value the gas density near $r \sim 1$ is $n \approx \ 10^9 - 10^{10}
\cc$ (since $v_r \approx c$ near the horizon).  The corresponding
magnetic field strength, assuming rough equipartition with the nearly
relativistic protons, is $B \approx 2 \times 10^3 \ $G.  At such
magnetic field strengths, relativistic electrons cooling by
synchrotron radiation would have a cooling time much less than the
inflow time of the gas.  In order to not overproduce the observed
radio to sub-mm luminosity of Sgr A$^*$, the bulk of the electrons
must therefore be marginally relativistic, with $T_e \sim
10^9-10^{10}$ K.  These plasma parameters ($n, B, T_e$) describe Bondi
and ADAF models of Sgr A$^*$ (e.g., Melia 1992, 1994; Narayan, Yi, \&
Mahadevan 1995; Narayan et al. 1998).  In such models the electrons
are assumed to be adiabatically compressed from large radii in the
accretion flow, with virtually no additional turbulent heating.

QN showed that accretion at much less than the Bondi rate could also
produce the observed high frequency emission from Sgr A$^*$, provided
the electrons were much hotter than in standard ADAF models (see their
Table 2 and Fig. 8b).  A simple explanation for this result can be
obtained by applying the Burbidge (1958) estimate to Sgr A$^*$.  We
consider synchrotron emission from a sphere of radius $R$ containing
relativistic electrons with a temperature $k T_e = \gamma m_e c^2$.
We take the electron heating rate to be comparable to the net
turbulent (magnetic) heating rate.  As can be confirmed {\it a
posteriori}, the synchrotron cooling time is $\gg$ the inflow time of
the gas.  The electron energy density is then similar to the magnetic
energy density
\begin{equation}
n\gamma m_ec^2 \approx {B^2\over 8\pi }.
\end{equation}
The frequency of peak synchrotron emission and the synchrotron
luminosity are given by
\begin{equation}
\nu \approx 0.1\gamma ^2{eB\over m_ec}
\end{equation}
and
\begin{equation}
L \approx \sigma _T c B^2 \gamma ^2 R^3 n,
\end{equation}
where $\sigma _T$ is the Thomson cross section. 

We express $n$, $\gamma$, and $B$ in terms of $R$, $\nu$, and $L$ (see
also Falcke 1996; Beckert \& Duschl 1997)
\begin{equation}
\gamma \approx 3.2 \left( {m_e\over c}~{\nu ^4 R^3\over L}\right) ^{1/7} \approx \ 100, \label{gamma}
\end{equation}
\begin{equation}
n \approx {4\over \gamma ^5 \lambda ^2r_e} \approx 10^6~{\rm cm}^{-3}, \label{n}
\end{equation}
and \begin{equation} B \approx \sqrt{8\pi m_ec^2\gamma n} \ \approx \
45~{\rm G}, \label{B}
\end{equation}
where $\lambda =c/\nu$ is the wavelength and $r_e=e^2/(m_ec^2)$ is the
classical electron radius.  For the numerical estimates in equations
(\ref{gamma})-(\ref{B}), we have used the observed values for Sgr A$^*$.
The peak synchrotron frequency is at $\nu \approx 10^3$ GHz with a
luminosity of $L \approx 10^{36} \egs$ (e.g., Serabyn et al. 1997).
In spherical accretion models, this high frequency emission arises
from very close to the black hole, so we have taken $R \approx R_g
\approx 10^{12}$ cm.

Equation (\ref{n}) gives a density close to the black hole of $n \sim
10^6 \cc$; the implied accretion rate is then $\sim 10^{-8} \mpy$,
three to four orders of magnitude smaller than the Bondi value.



The thermal blackbody emission at frequency $\nu$ from a sphere of
radius $R$ is \beq L_t = 2 \pi \nu^3 \gamma m_e 4 \pi R^2 \approx
10^{37} \egs, \eeq where the numerical estimate is for our fiducial
parameters.  This comparison shows that the synchrotron emission
becomes optically thin below the peak frequency, near $\nu \approx
300$ GHz.  At lower frequencies the emission is self-absorbed.

The above considerations show that both low ($\sim 10^{-8} \mpy$) and
high ($\sim 10^{-5}-10^{-4} \mpy$) accretion rate models can explain
the observed sub-mm ``bump'' in Sgr A$^*$.  Such models can be
distinguished by comparing the observed brightness temperature and/or
radio image as a function of frequency with the theoretical
predictions (see, e.g., \"Ozel, Psaltis, \& Narayan 2000).  This test
has been difficult to implement because interstellar scattering
significantly broadens the image of Sgr A$^*$.\footnote{Recent
detections of Sgr A$^*$'s intrinsic size (Lo et al. 1998; Krichbaum et
al. 1998) still have sufficient uncertainties that a range of
theoretical models are allowed.} In the next section we show that the
linear polarization of Sgr A$^*$ at high frequencies provides an
additional discriminant.


\section{Faraday Depolarization}

The anisotropic index of refraction of a magnetized plasma leads to a
frequency-dependent rotation in the position angle, $\theta$, of
linearly polarized electromagnetic waves, \beq \theta = RM \lambda^2,
\eeq where $RM$ is the rotation measure.  This can lead to significant
depolarization of intrinsically linearly polarized synchrotron
emission.

For a ``cold'' non-relativistic plasma, RM is given by (e.g., Rybicki
\& Lightman 1979) \begin{eqnarray} RM &=&{e^3\over 2\pi m_e^2c^4}\int
d{\bf l}\cdot {\bf B}n \nonumber \\ &=& 2.63\times 10^{-13}\times \int
d{\bf l}\cdot {\bf B}n~{{\rm rad}\over {\rm m}^2}, \label{cold1}
\end{eqnarray}
where $d {\bf l}$ is the differential path length from the observer to
the source.  In the Appendix we show that the rotation measure for an
ultrarelativistic thermal plasma is given by
\begin{eqnarray}
RM_{\gamma }&=&{e^3\over 2\pi m_e^2c^4}\int d{\bf l}\cdot {\bf B}n {\log
\gamma \over 2\gamma ^2}\nonumber \\ &=& 2.63\times 10^{-13}\times \int d{\bf
l}\cdot {\bf B}n{\log \gamma \over 2\gamma ^2}~{{\rm rad}\over {\rm
m}^2},
\label{ultra1}
\end{eqnarray}
where $\gamma = kT_e/m_ec^2$.  A comparable expression is obtained for
a power law distribution of relativistic electrons, with $\gamma$
replaced by $\gamma_{min}$, the minimum Lorentz factor of the
electrons (Jones \& O'Dell 1977).  In what follows, we define $RM(r)$
to be the contribution to the net rotation measure from radii within
$dr \approx r$ of radius $r$ in the accretion flow.

\subsection{ADAF/Bondi Models}

In spherical accretion flow models, higher frequency radio emission
arises from closer to the black hole, where the electron temperature
and magnetic field strengths are the largest; this is also true for
jet models (e.g., Falcke 1999).  \"Ozel et al. (2000) show that in
ADAF models of Sgr A$^*$ the synchrotron emission at frequency $\nu =
100 \ \nu_{100}$ GHz arises from a radius $r_\nu \approx 20 \
\nu_{100}^{-0.9}$ (see their Fig. 5).\footnote{Falcke (1999) finds a
similar expression for $r_\nu$ in the jet model.}  This radius defines
the $\tau = 1$ surface of the synchrotron emission.  For smaller radii
the emission is self-absorbed while for larger radii it is optically
thin.  Faraday rotation is only important for $r \gsim r_\nu$, where
the photons ``free stream'' out of the accretion flow.

In Bondi/ADAF models Faraday rotation is so strong that the
synchrotron emission is completely depolarized by the plasma within
$dr \sim r$ of $r_\nu$, i.e., in the vicinity of the $\tau = 1$
surface where it is emitted (depolarization and emission are thus
virtually co-spatial).  Taking $n \propto r^{-3/2}$ and $B \propto
r^{-5/4}$; the rotation measure scales roughly as $RM \propto r^{-7/4}
\gamma^{-2}$.  The relativistic suppression of the rotation measure is
small in all models which have an accretion rate comparable to the
Bondi rate, because the electrons must then be at most marginally
relativistic (\S2).  The rotation measure as a function of radius is
thus given by (see also Bower et al. 1999ab) \beq RM \approx 10^{13}
r^{-7/4} \rad. \label{ADAF} \eeq In equation (\ref{ADAF}) we have
assumed that the magnetic field is in rough equipartition with the gas
pressure, has a significant component along the line of sight, and has
a coherence length $\ell \sim r$; for $\ell \ll r$, $RM$ is reduced by
$\approx (\ell/r)^{1/2}$.  The normalization in equation (\ref{ADAF})
is set by the Bondi accretion rate.

The large $RM$ in ADAF/Bondi models leads to a significant rotation in
the position angle of linearly polarized waves.  Photons of frequency
$\nu$ emitted at radius $r_\nu$ undergo Faraday rotation through an
angle \beq \theta_\nu \approx \lambda^2 RM(r_\nu) \approx 10^8
\nu_{100}^{-2} r_\nu^{-7/4} \sim \ 10^6 \nu_{100}^{-0.43} \ {\rm rad},
\eeq where the last estimate uses \"Ozel et al.'s (2000) fit to
$r_\nu(\nu)$.

These rotation angles are so large that the synchrotron emission in
ADAF/Bondi models of Sgr A$^*$ is completely depolarized by Faraday
rotation. For example, in a simple uniform source model, the observed
polarization is $\propto \theta_\nu^{-1}$ (Pacholczyk 1970).  In
general, the observed polarization depends on the rotation measure
power spectrum, but is $\ll 1$ for $\theta_\nu \gg 1$ (e.g., Tribble
1991).\footnote{One way of evading this conclusion is to posit that
the magnetic field is sufficiently tangled ($\ell \ll r$) to decrease
$RM$ to $\lsim 10^{6} \rad$.  This tangling would, however, also
eliminate linear polarization.}

\subsection{$\dot M \ll \dot M_{\rm Bondi}$}

If the accretion rate onto Sgr A$^*$ is much less than the Bondi rate,
significant polarization may be observable at high frequencies; we
show this using an order of magnitude estimate.

For the $\dot M \sim 10^{-8} \ \mpy$ model of \S2 the rotation measure
calculated using equation (\ref{ultra1}) is $RM \approx 10^3 \rad$
near $r \sim 1$.  Moreover, if $n \propto r^{-1/2}$, as in CDAF
models, the magnetic field scales as $B \propto r^{-3/4}$ and \beq RM
\approx 10^3 r^{-1/4} \left(\gamma \over 100\right)^{-2}
\rad. \label{CDAF} \eeq The variation of the electron Lorentz factor
with radius is somewhat uncertain, but we expect roughly $\gamma
\propto r^{-1}$, so that the electrons become non-relativistic by $r
\sim 10^2$.  Equation (\ref{CDAF}) then shows that RM has its maximal
value at large radii, $r \sim 10^2$, where $RM \sim 3 \times 10^6
\rad$.


Equation (\ref{CDAF}) demonstrates that, in contrast to ADAF models,
there is no depolarization of synchrotron emission at small radii in
models with accretion rates much less than the Bondi rate; $RM$ is
negligible in the region where the synchrotron emission is produced.
Depolarization can still be important, however, because observed
photons experience different Faraday rotation at large radii, $r \gsim
10^2$, on their way out of the accretion flow (e.g., Bower et
al. 1999ab).


Spatial variation in the rotation measure will depolarize Sgr A$^*$ at
frequencies for which $\delta \theta = \lambda^2 \delta RM \gsim \pi$,
i.e., for \beq \nu \lsim 100 \left({\delta RM \over 10^6
\rad}\right)^{1/2} {\rm GHz}, \label{numax}\eeq where $\delta RM$ is
the difference in the rotation measure for photons of a given
frequency which travel through different parts of the accretion
flow.\footnote{To be precise, $\delta RM$ is the difference in the
rotation measure at $r \sim 100-10^4$ on scales of $r_\nu$, the source
size at frequency $\nu$.  This is difficult to calculate analytically,
but could be determined from future MHD simulations of non-radiating
accretion flows.} Quantitative calculations of depolarization by
differential Faraday rotation are uncertain; two points are, however,
clear: (1) At low frequencies, $\ll 100$ GHz, Sgr A$^*$ is easily
depolarized at $r \gsim 10^2$.  The required $\delta RM$ is $\ll 10^6
\rad$, orders of magnitudes smaller than the values of $RM$ obtained
at $r \sim 10^2-10^4$.  (2) Emission above $\sim 100$ GHz can
plausibly be linearly polarized if the accretion rate onto Sgr A$^*$
is much less than the Bondi rate.  In particular, equations
(\ref{CDAF}) and (\ref{numax}) show that for $\dot M \ll \dot M_{\rm
Bondi}$, emission above $\approx 100$ GHz is not depolarized
propagating out of the accretion flow.

\section{Discussion}

ADAF/Bondi models assume that the accretion rate onto Sgr A$^*$ is of
order the Bondi rate ($\sim 10^{-4}-10^{-5} \mpy$) and that the radio
to infrared emission is produced by synchrotron emission from
marginally relativistic electrons ($T_e \approx 10^9-10^{10}$ K).  In
such models the rotation measure is $\gsim 10^{10} \rad$ inside
$\approx 100$ Schwarzschild radii where the synchrotron emission is
produced.  ADAF/Bondi models thus predict that Sgr A$^*$ should be
depolarized by Faraday rotation over the entire radio to infrared
spectrum, and should have nearly zero linear polarization.

The theoretical arguments summarized in \S1 propose that the accretion
rate onto Sgr A$^*$ is much less than the Bondi rate.  We have
described one such model, in which the electron heating rate is of
order the rate of change of the magnetic energy density.  For an
accretion rate $\sim 10^{3}$ times smaller than the Bondi rate, i.e.,
$\sim 10^{-8} \mpy$, and with relativistic electrons with $\gamma
\approx 100$, this model can explain the observed high frequency
emission from Sgr A$^*$.  Moreover, it predicts that the rotation
measure in the accretion flow is much smaller than in ADAF/Bondi
models. This is because the gas density and magnetic field strength
close to the black hole are much smaller, and because the electrons
are relativistic ($RM \propto \gamma^{-2} \log\gamma$ for $\gamma \gg
1$; see \S3 and the Appendix).  The maximal contribution to the
rotation measure comes from $\sim 10^2-10^3$ Schwarzschild radii,
where $RM \sim 10^6 \rad$.

Rotation measures of $\sim 10^6 \rad$ can depolarize Sgr A$^*$ at $\nu
\ll 100$ GHz by differential Faraday rotation; photons of a given
frequency travel through different rotation measures on their way out
of the accretion flow.  Following Bower et al. (1999ab), we believe
that this accounts for the $\lsim 0.2 \%$ linear polarization of Sgr
A$^*$ at low frequencies (from $\approx 4$ to $\approx 23$ GHz; see
Bower et al. 1999ab);\footnote{Bower et al. (1999ab) showed that
``bandwidth'' depolarization at $\nu \approx 8$ GHz requires $RM >
10^7 \rad$; this constraint does not, however, apply to the
depolarization discussed here, namely that due to a spatially varying
$RM$.} it is less clear, however, that it can account for Bower et
al.'s (1999b) limit of $\lsim 1 \%$ linear polarization at $86$ GHz
(see below).  In fact, rotation measures of $\approx 10^6 \rad$ are
insufficient to depolarize emission above $\approx 100$ GHz.  As a
result, in models with accretion rates much less than the Bondi rate,
$\gsim 100$ GHz emission is not depolarized propagating out of the
accretion flow; intrinsically polarized synchrotron emission may
therefore be observable at high frequencies.

The above considerations show that the linear polarization of Sgr
A$^*$ at high frequencies provides a means of distinguishing between
accretion at the Bondi rate, and accretion at a much smaller rate.  In
fact, Aitken et al. (2000) report a possible detection of $\sim 10 \%$
linear polarization from Sgr A$^*$ between $150$ and $400$ GHz.  If
confirmed, these observations require an accretion rate onto Sgr A*
much less than the Bondi rate, roughly $\dot M \sim 10^{-8} \mpy$.

One difficulty in interpreting Aitken et al's results is the large
beam ($\approx 20''$) of the SCUBA camera on the JCMT.  This large
beam forced Aitken et al. to subtract out free-free and (polarized!)
dust emission in order to isolate the flux and polarization of Sgr
A$^*$.  Future high resolution polarimetry at mm wavelengths is
clearly necessary to further address this important issue.

Aitken et al. find that the position angle of Sgr A$^*$ changes by
$\lsim 10^o$ between $\lambda = 0.135$ cm and $\lambda = 0.2$ cm; at
face value this implies $RM \lsim 10^5 \rad$, somewhat smaller than
the values of $\sim 10^6 \rad$ in our model.  This assumes, however,
that the intrinsic position angle of Sgr A$^*$ is the same at $\lambda
= 0.135$ cm and $\lambda = 0.2$ cm, which need not be the case.
Moreover, our estimates of $RM$ are actually upper limits, since they
assume (1) equipartition magnetic fields aligned along the line of
sight and (2) that our line of sight passes through the equatorial
plane of the accretion flow.



Our analysis of depolarization is applicable even if the radio
emission from Sgr A$^*$ is dominated by a jet/outflow, rather than the
accretion flow as we have assumed.  In jet models, it is still natural
for the highest frequency emission to originate very close to the
black hole; in Falcke's model, e.g., the $\gsim 100$ GHz emission
arises from $\lsim 10 \ R_g$, in what is really a ``transition
region'' between the accretion flow and the jet (Falcke 1999).  In
order for this emission to not be depolarized (either {\it in situ} or
propagating through the accretion flow), our constraints on the
rotation measure and the plasma conditions close to the black hole
still apply.

Two scenarios in which accretion at the Bondi rate could be consistent
with observed linear polarization at high frequencies are (1) if the
high frequency emission arises close to the black hole, but in a
nearly empty funnel pointed directly towards us (e.g., along the
rotation axis of an ADAF) or (2) if the high frequency emission from
Sgr A$^*$ is produced at very large distances from the black hole, $r
\gsim 10^3$.  The former possibility requires a rather special
geometry\footnote{For example, in Stone et al.'s (1999) simulations of
non-radiating accretion flows, the density varies with polar angle
roughly as $\rho \propto \sin^2 \theta$ (see also Quataert \& Gruzinov
2000); thus for $\dot M \sim \dot M_{\rm Bondi}$, the emission must be
confined to $\theta \lsim {\rm 3}^o$ in order for the density to be
sufficiently small that high frequency emission is not depolarized.
In addition our line of sight must lie within $\lsim 3^o$ of the
rotation axis of the flow.} and the latter is ruled out by the VLBI
source size of $\sim 10 \ R_g$ (Krichbaum et al. 1998) and the
variability of Sgr A$^*$ at $\approx 100$ GHz (Tsuboi, Miyazaki, \&
Tsutsumi 1999).

\subsection{Application to Other Systems}

Although we have have focused our analysis on Sgr A$^*$ at the
Galactic Center, linear polarization of high frequency radio emission
can be used as a probe of the accretion physics in other
low-luminosity galactic nuclei (see, e.g., Nagar et al. 2000 for
recent VLA observations of LLAGN).  For a black hole of mass $M = m_9
10^9 M_\odot$ accreting (spherically) at a rate $\dot M = 10^{-4} \md
\dot M_{\rm edd} \approx 10^{23} \md m_9$ g s$^{-1}$, the density,
magnetic field strength, and rotation measure in ADAF models are \beq
n \approx 3 \times 10^{6} \ \md \ m_9^{-1} \ r^{-3/2} \cc,
\label{n} \eeq \beq B \approx 100 \ \md^{1/2} \ m_9^{-1/2} \ r^{-5/4} \ {\rm
G}, \label{B2} \eeq and \beq RM \approx 3 \times 10^{10} \ \md^{3/2} \
m_9^{-1/2} \ r^{-7/4} \ \rad. \label{RMf} \eeq Equation (\ref{RMf})
shows that large rotation measures and the associated depolarization
of synchrotron emission by Faraday rotation are generic features of
ADAF models (unless $\md \ll 1$).

The absence of observed linear polarization in the radio spectrum of a
low-luminosity galactic nucleus would be consistent with ADAF models.
By contrast, detected linear polarization would argue against an ADAF
as the source of the observed radio emission.  A particularly
interesting class of systems for future polarimetry are elliptical
galaxies in nearby X-ray clusters (e.g., NGC 4649, 4472, and 4636 in
the Virgo cluster).  As discussed by, e.g., Fabian \& Canizares
(1988), Fabian \& Rees (1995), and Di Matteo et al. (1999, 2000), many
of these galaxies have extremely dim nuclei given the inferred black
hole masses ($\sim 10^9 M_\odot$) and Bondi accretion rates.  Linear
polarization may shed important light on the physics of these systems.


\acknowledgments We thank Don Backer, Roger Blandford, and Mark Reid
for useful correspondence, Bruce Draine for useful conversations, and
John Bahcall, Heino Falcke, and Feryal \"Ozel for helpful comments on
the paper.  EQ is supported by NASA through Chandra Fellowship
PF9-10008, awarded by the Chandra X--ray Center, which is operated by
the Smithsonian Astrophysical Observatory for NASA under contract NAS
8-39073. AG was supported by the W. M. Keck Foundation and NSF
PHY-9513835.

\newpage

\begin{appendix}

\section{Faraday rotation in an ultra-relativistic Maxwellian plasma}

Faraday rotation in a cold plasma is described by a change in position
angle given by
\begin{equation}
{d \theta \over dl}={k_{\parallel }\over 2}{\omega _p^2\omega _B\over
\omega ^3},
\end{equation}
where $\omega =ck$ is the frequency of the radio wave, $k_{\parallel
}$ is the projection of the wavenumber along the magnetic field,
$\omega _p^2=4\pi ne^2/m_e$ is the plasma frequency, and $\omega
_B=eB/(m_ec)$ is the cyclotron frequency. This corresponds to the
usual rotation measure
\begin{equation}
RM\equiv {\theta \over \lambda ^2}={e^3\over 2\pi m_e^2c^4}\int d{\bf
l}\cdot {\bf B}n ~=~2.63\times 10^{-13}\times \int d{\bf l}\cdot {\bf
B}n~{{\rm rad}\over {\rm m}^2}. \label{cold}
\end{equation}
Here we derive the rotation measure for an ultrarelativistic
Maxwellian plasma:
\begin{equation}
RM_{\gamma }={e^3\over 2\pi m_e^2c^4}\int d{\bf l}\cdot {\bf B}n {\log \gamma \over 2\gamma ^2}~=~2.63\times 10^{-13}\times \int d{\bf l}\cdot {\bf B}n{\log \gamma \over 2\gamma ^2}~{{\rm rad}\over {\rm m}^2},
\label{ultra}
\end{equation}
where we have defined $\gamma \equiv k T_e/(m_ec^2)$.  The dominant
correction to the non-relativistic expression is the relativistic
mass: $m_e \rightarrow \gamma m_e$.

We use the Vlasov equations to calculate the plasma permittivity and
hence the dispersion relation for electromagnetic waves. For a
magnetic field and wavenumber along the z axis, the first-order (in
the unperturbed magnetic field) permittivity is given by
\begin{equation}
\epsilon ^{(1)}_{xy}~=~{-i\over 2\omega} {4\pi e^2\over m_e} {eB\over m_ec} \int d^3p{1\over (\omega -kv_z)^2}{p_{\perp }^2\over p}{dF\over dp}{m_e^2c^2\over p^2+m_e^2c^2},
\label{perm}
\end{equation}
where the unperturbed distribution function is normalized by $\int d^3p F(p)=n$, and $p_\perp^2\equiv p_x^2+p_y^2$. For a cold plasma, equation (\ref{perm}) gives 
\begin{equation}
\epsilon _{xy}={i\omega _p^2\omega _B\over \omega ^3}.
\end{equation}
Using standard arguments (e.g., Rybicki \& Lightman 1979) this leads
to the RM for a cold plasma given by equation (\ref{cold}). For an
ultra-relativistic plasma, equation (\ref{perm}) gives
\begin{equation}
\epsilon _{xy}={i\omega _p^2\omega _B\over \omega ^3}{\log \gamma
\over 2\gamma ^2},
\label{ultraperm}
\end{equation}
where we have not changed the definition of $\omega _p$ and $\omega
_B$ in the ultra-relativistic regime.  Equation (\ref{ultraperm}) for
the permittivity gives the ultra-relativistic RM in equation
(\ref{ultra}).

\end{appendix}

\end{document}